# Nonlinear viscoelastic wave propagation: an extension of Nearly Constant Attenuation (NCQ) models.

Nicolas Delépine[1], Luca Lenti[2], Guy Bonnet[3], Jean-François Semblat, A.M.ASCE[4]

**Subject headings:** Inelasticity; Viscoelasticity; Damping; Wave propagation; Earthquake engineering; Ground motion; Nonlinear response; Finite element method.

## Abstract

Hysteretic damping is often modeled by means of linear viscoelastic approaches such as "nearly constant Attenuation (NCQ)" models. These models do not take into account nonlinear effects either on the stiffness or on the damping, which are well known features of soil dynamic behavior. The aim of this paper is to propose a mechanical model involving nonlinear viscoelastic behavior for isotropic materials. This model simultaneously takes into account nonlinear elasticity and nonlinear damping. On the one hand, the shear modulus is a function of the excitation level; on the other, the description of viscosity is based on a generalized Maxwell body involving non-linearity. This formulation is implemented into a 1D finite element approach for a dry soil. The validation of the model shows its ability to retrieve low amplitude ground motion response. For larger excitation levels, the analysis of seismic wave propagation in a nonlinear soil layer over an elastic bedrock leads to results which are physically satisfactory (lower amplitudes, larger time delays, higher frequency content).

## 1 Introduction

The analysis of seismic wave propagation in alluvial basins is complex since various phenomena are involved at different scales (Semblat and Pecker, 2009): resonance at the scale of the whole basin (Bard and Bouchon, 1985; Paoluci, 1999; Semblat et al., 2003), surface waves generation at the basin edges (Bard and Riepl-Thomas, 2000; Bozzano et al., 2008; Kawase, 2003; Moeen-Vaziri and Trifunac, 1988; Semblat et al., 2000, 2005; Sánchez-Sesma and Luzón, 1995), soil nonlinear behavior at the geotechnical scale (Bonilla et al., 2006; Iai et al., 1995; Kramer, 1996). Handling these different features of seismic wave propagation at the same time may be important because the interaction between, for instance, surface wave generation and shear modulus degradation may be significant. The impact on the amplification process could thus be very large and complex.

Nonlinear constitutive equations are very important in the case of strong ground motion since the mechanical behavior of many soils depends on the excitation level and on the loading history. In this work, the attention is focused on the aspects of nonlinear behavior of dry isotropic soils submitted to dynamic loadings. Various approaches are available to model the dependence of the mechanical features of soils on the excitation level: equivalent linear model and nonlinear cyclic constitutive equations (including plasticity).

The equivalent linear model approximates the problem in the linear range using an iterative procedure (Schnabel et al., 1972). Since this model leads to over-damped higher frequency components, recent researches improved it by introducing both frequency or mean stress dependencies of the soil properties (Sugito, 1995; Kausel and Assimaki, 2002). Several

[1] Université Paris-Est, LCPC, presently at: IFP, 1 & 4 av. de Bois-Préau, 92852 Rueil-Malmaison Cedex, France, nicolas.delepine@ifp.fr
[2] Université Paris-Est, LCPC, 58 bd Lefebvre, 75732 Paris Cedex 15, France, lenti@lcpc.fr
[3] Université Paris-Est, Champs sur Marne, France, bonnet@univ-paris-est.fr
[4] Université Paris-Est, LCPC, 58 bd Lefebvre, 75732 Paris Cedex 15, France, semblat@lcpc.fr (correspond. author)

comparisons involving such models were proposed by Bonilla et al. (2006) and Kwok et al. (2008).

Concerning nonlinearity, some models are based on both the "hyperbolic law", for describing shear modulus reduction curves, and on the Masing criterion (Masing, 1926) for the description of unloading and reloading phases. Such models have been widely developed (Matasovic, 1993; Matasovic and Vucetic, 1995). However, these models generally need large computational efforts and often lack of a strong mechanical basis, e.g. thermodynamics (Lemaître and Chaboche, 1992). Some other models are fully elastoplastic (Aubry et al., 1982; Prevost, 1985; Gyebi and Dasgupta, 1992) or include dependence on confining pressure (Hashash and Park, 2001; Park and Hashash, 2004) and pore pressure (Bonilla et al., 2005). However, their use for large scale wave propagation analyses is limited as a consequence of the large number of parameters needed and the frequency/wavelength range to investigate.

In this paper, a 3D nonlinear viscoelastic model is proposed. This model simultaneously follows a nonlinear elastic law and a nonlinear viscous law to investigate the ground response to strong seismic excitation.

## 2 Mechanical formulation of the model

### 2.1 3D linear viscoelasticity

#### 2.1.1 General formulation

The 3D formulation of the viscoelastic model starts from the following relation:

$$\sigma_{ij} = s_{ij} + p\delta_{ij} \tag{1}$$

where $\sigma_{ij}$, $s_{ij}$, $\delta_{ij}$ and $p$ are the Cauchy stress tensor, the deviatoric stress tensor, the Kronecker unit tensor and the volumetric tension respectively. For an isotropic material, we can write:

$$p = K \cdot e_{kk} \tag{2}$$

where $K$ and $e_{kk}$ are the bulk modulus and the volumetric strain respectively. The relation between the components of the deviatoric stress tensor $s$ and the shear deviatoric strain tensor $e$ in the case of linear viscoelasticity is formulated in the frequency domain as simply as:

$$s_{ij}(\omega) = 2M(\omega)e_{ij}(\omega) \tag{3}$$

$s_{ij}(\omega)$, $e_{ij}(\omega)$ are the Fourier transforms of the components of the deviatoric stress and strain tensors. $M(\omega)$ is the complex-valued, frequency-dependent, viscoelastic modulus from which we can define the specific attenuation $Q^{-1}$ in the following way (Bourbié et al., 1987; Semblat and Pecker, 2009):

$$2\xi = Q^{-1}(\omega) \approx Im(M(\omega))/Re(M(\omega)) \tag{4}$$

where $\xi$ is the damping ratio and $Re$ and $Im$ are the real and imaginary parts of a complex variable (resp.).

#### 2.1.2 NCQ models

This family of models is defined in term of the quality factor $Q$. A nearly constant $Q$ in a broad frequency range and for a given strain level is introduced. Biot (1958) first demonstrated that a causal form of hysteretic damping can be simulated by viscoelastic cells in parallel. Liu et al. (1976) constructed such models by direct superposition of Zener cells (standard solid). Emmerich and Korn (1987) improved and extended the Padé approximation (Day and Minster, 1984) by considering a generalized $n$-cells Maxwell body (Fig. 1, left). Mozco and Kristek (2005) proved the equivalence of the models of Liu and Emmerich and Korn. The implementation proposed by Emmerich and Korn is used in the following.

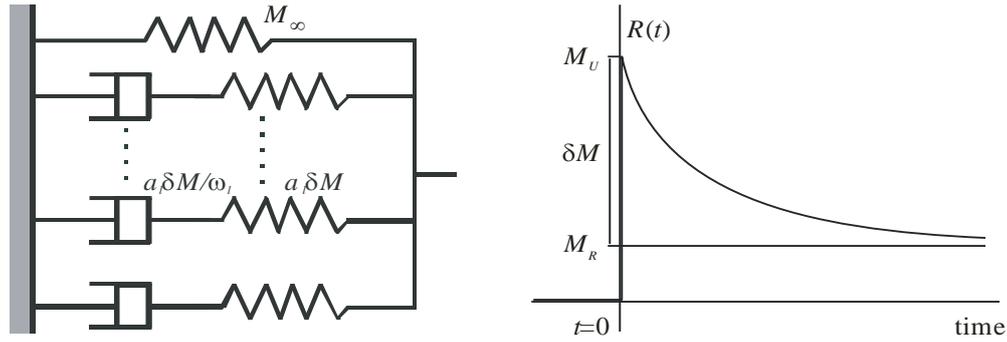

**Fig. 1.** Generalized Maxwell body with viscosities $a_l.\delta M / \omega_l$ and elastic moduli $a_l.\delta M$ for each rheological cell (left). Typical relaxation function $R(t)$ (right) with $M_U$ the unrelaxed modulus and $M_R=M_U-\delta M$ the relaxed modulus.

The generalized Maxwell model leads to the frequency dependent complex modulus (variables with bracket are not tensorial):

$$M(\omega) = M_U \left( 1 - \frac{\sum_{l=1}^{n} y_{(l,0)} \omega_{(l)} /(i\omega + \omega_{(l)})}{1 + \sum_{l=1}^{n} y_{(l,0)}} \right) \quad (5)$$

$M_U$ is the unrelaxed (instantaneous) modulus and $M_R$ is the relaxed (long term) modulus (Fig. 1, right). The $y_{(l,0)}$ variables characterize the rheological model and are calculated by means of an optimization method in order to obtain a nearly constant attenuation in a given frequency range (see Appendix).

Using Eqs. **(4)** and **(5)**, the quality factor has the following expression:

$$Q^{-1}(\omega) = \frac{\operatorname{Im} M(\omega)}{\operatorname{Re} M(\omega)} = \frac{\sum_{l=1}^{n} y_{(l,0)} \frac{\omega / \omega_{(l)}}{1 + (\omega / \omega_{(l)})^2}}{1 + \sum_{l=1}^{n} y_{(l,0)} \frac{(\omega / \omega_{(l)})^2}{1 + (\omega / \omega_{(l)})^2}} \quad (6)$$

The $\omega_{(l)}$ frequencies characterize each individual rheological cell (see Appendix).

The constitutive equations for the linear viscoelastic model are thus:

$$s_{ij}(t) = 2M_U \left[ e_{ij}(t) - \sum_{l=1}^{n} \zeta_{(l)}(t) \right] \quad (7)$$

and

$$\dot{\zeta}_{(l)}(t) + \omega_{(l)} \zeta_{(l)}(t) = \omega_{(l)} \frac{y_{(l,0)}}{1 + \sum_{l=1}^{n} y_{(l,0)}} e_{ij}(t) \quad (8)$$

where $\zeta_{(l)}(t)$ are relaxation parameters physically related to the anelastic deformation of the $l^{th}$-cell (Fig. 1, left).

Fig. 2 displays the attenuation curve (a), $2\xi=Q^{-1}$, and the phase velocity (b), $V_{ph}$, as functions of frequency. These graphs are obtained considering 3 Zener's cells which are equivalent to generalized Maxwell cells (Fig. 1, left). The attenuation is nearly constant, $2\xi=Q^{-1}=0.05$, in the frequency range 0.1-10Hz.

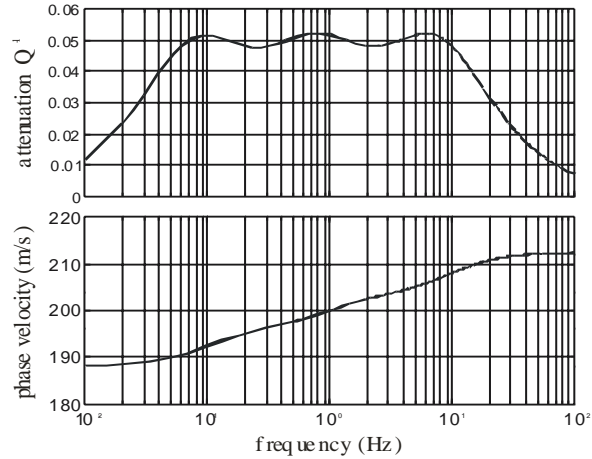

**Fig. 2.** Generalized Maxwell body approximating a Nearly Constant Quality factor Q≈20 (top), in the frequency range 0.1-10Hz, and corresponding phase velocity $V_{ph}$ (bottom). The target phase velocity, $|V_{ph}|$=200m/s, is chosen at a frequency of 1Hz.

### 2.2 3D nonlinear viscoelastic model

#### 2.2.1 *Principles of the nonlinear model*

In order to describe the soil's shear modulus and damping variations with the excitation level, an elastic potential function and a dissipation function depending on the magnitude of the second invariant of the strain tensor are introduced. The description of viscosity is based on a Nearly Constant Attenuation model able to fulfil the causality principle for seismic wave propagation (dispersive materials). Owing to the frequent use of this model within the geophysical community, it is usually called Nearly Constant Quality Factor (or "NCQ") model. At the same time, it leads to a constant value of the damping factor at low strains over a broad frequency range of engineering interest (Kjartansson, 1979). The model is well-adapted to time domain formulations (some alternative numerical strategies are available (Carcione et al., 2002; Munjiza et al., 1998; Semblat, 1997)).

In the NCQ model, we introduce a dependence on the excitation level in order to consider an increasing damping ratio suggested from earthquakes records and geotechnical data (Iai et al., 1995; Vucetic, 1990). This dependence is controlled during the 3D stress-strain path by the variation of the second order invariant of the strain tensor.

#### 2.2.2 *Formulation of the extended NCQ model ("X-NCQ")*

To account for non linear behavior of soils in the case of any 3D stress-strain path, Eq. **(7)** is extended as follows:

$$s_{ij}(t) = 2M_U(J_2)\left[e_{ij}(t) - \sum_{l=1}^{n} \zeta_{(l)}(t, y_{(l)}(J_2))\right] \quad (9)$$

where $J_2$ is the second invariant of the deviatoric strain tensor, defined from the following relations:

$$J_2 = I_2' - \frac{I_1'^2}{3} \quad (10)$$

with the 2 first invariants of the strain tensor:

$$I_1' = trace(\varepsilon) \quad (11)$$

and

$$I_2' = \frac{1}{2} \cdot trace(\varepsilon^2) \tag{12}$$

In addition, the shear modulus is assumed to change during the global stress-strain path according to the following relation:

$$M_U(J_2) = M_{U,0}[1 - \Phi(J_2)] \tag{13}$$

with

$$\Phi(J_2) = \frac{\alpha |J_2|^{\frac{1}{2}}}{1 + \alpha |J_2|^{\frac{1}{2}}} \tag{14}$$

and where $M_{U,0}$ denotes the unrelaxed modulus characterizing the instantaneous response of the soil at small strains and $\alpha$ is a parameter quantifying its nonlinear behavior for larger strains.

The octahedral strain $\gamma_{oct}$ is now introduced:

$$|\gamma_{oct}| = 2|J_2|^{\frac{1}{2}} \tag{15}$$

It leads to

$$M_U(|\gamma_{oct}|) = M_{U,0}[1 - \Phi(|\gamma_{oct}|)] \tag{16}$$

where:

$$\Phi(|\gamma_{oct}|) = \frac{\alpha |\gamma_{oct}|/2}{1 + \alpha |\gamma_{oct}|/2} \tag{17}$$

Such a dependence of the nonlinear elastic modulus on the octahedral strain also implies a strain dependence for the variables $y_{(l)}$ and $\zeta_{(l)}$. Determination of damping ratio $\xi$ has been performed by Strick (1967) using wave propagation measurements. Formulations for the dependence of the damping ratio $\xi$ on the shear strain modulus have been proposed by Hardin and Drnevich (1972). In the case of 3D loadings, different authors (El Hosri et al., 1984; Heitz, 1992; Bonnet and Heitz, 1994) proposed an extension of $\xi$, such as:

$$\xi(|\gamma_{oct}|) = \xi_0 + (\xi_{max} - \xi_0)\Phi(|\gamma_{oct}|) \tag{18}$$

where $\xi_0$ and $\xi_{max}$ characterize the dissipated energy in the small and larger strain ranges respectively. Typical $M_U(\gamma)=G(\gamma)$ and $\xi(\gamma)$ curves are proposed in Fig. 3.

The damping ratio $\xi$ and the attenuation $Q^{-1}$ are now related by:

$$Q^{-1} = 2\xi(|\gamma_{oct}|) \tag{19}$$

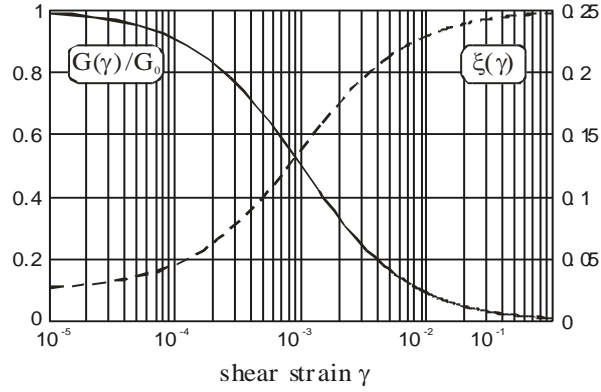

**Fig. 3.** Typical nonlinear dynamic properties of soils: shear modulus reduction (solid) and damping increase (dashed) with increasing shear strain.

### 2.2.3  Features of the extended NCQ model

In paragraph 2.1.1, the solution of Eq. (**9**) in the limit of low excitation levels has been found. For low octahedral strains, we can consider that:

$$Q_0^{-1} = 2\xi_0 \qquad (20)$$

and

$$M_U(|\gamma_{oct}| \approx 0) = M_{U,0} = G_0 \qquad (21)$$

For every other value of the induced strain, the $Q^{-1}$ factor increases with strain according to Eq. (**19**). This change has no influence on the frequency range in which $Q^{-1}$ is constant. In other words, in Eq. (**6**) only the variables $y_{(l,0)}$ change to account for the variation of the damping with strain. We therefore introduce a strain variation of the variables $y_{(l)}$ with strain in the following form:

$$y_{(l)}(|\gamma_{oct}|) = c(|\gamma_{oct}|) y_{(l,0)} \qquad (22)$$

Using Eqs. (**4**), (**15**) and (**17**), for every level of induced octahedral strain, Eqs. (**5**) and (**34**) can be rewritten in the following form, respectively:

$$M(\omega, |\gamma_{oct}|) = M_U(|\gamma_{oct}|) \left( 1 - \frac{c(|\gamma_{oct}|) \sum_{l=1}^{n} y_{(l,0)} \omega_{(l)} / (i\omega + \omega_{(l)})}{1 + c(|\gamma_{oct}|) \sum_{l=1}^{n} y_{(l,0)}} \right) \qquad (23)$$

$$Q^{-1}(\omega, |\gamma_{oct}|) \approx c(|\gamma_{oct}|) \sum_{l=1}^{n} y_{(l,0)} \frac{\omega/\omega_{(l)}}{1 + (\omega/\omega_{(l)})^2} \qquad (24)$$

where, using Eq. (**18**), $c(|\gamma_{oct}|)$ is given by:

$$c(|\gamma_{oct}|) = \frac{Q^{-1}(\omega, |\gamma_{oct}|)}{Q_0^{-1}} = \frac{\xi(|\gamma_{oct}|)}{\xi_0} = \left[ 1 + \frac{\xi_{max} - \xi_0}{\xi_0} \Phi(|\gamma_{oct}|) \right] \qquad (25)$$

For every level of induced octahedral strain, Eq. (**8**) can be written in the more general form:

$$\dot{\zeta}_{(l)}(t) + \omega_{(l)} \zeta_{(l)}(t) = \omega_{(l)} \frac{c(|\gamma_{oct}|) y_{(l,0)}}{1 + c(|\gamma_{oct}|) \sum_{l=1}^{n} y_{(l,0)}} e_{ij}(t) \qquad (26)$$

The latter expression and Eq. (**16**) are used to solve Eq. (**9**) in the time domain.

### 2.3 Synthesis: 1D case

For a unidirectional propagating shear wave, $|\gamma_{oct}|$ is equal to $2|\gamma|$, where $\gamma$ is the shear strain. Equation **(16)** can be written in the form:

$$M_U(|\gamma|) = G(|\gamma|) = \frac{G_0}{1+\alpha|\gamma|} \tag{27}$$

In this case, Eq. **(27)** expresses a hyperbolic law for the reduction of the shear modulus as the one proposed by Hardin and Drnevich (1972). As a consequence, the following equation for the function $c(|\gamma_{oct}|)$ is obtained:

$$c(|\gamma|) = \left[1 + \frac{\xi_{max} - \xi_0}{\xi_0}\left(\frac{\alpha|\gamma|}{1+\alpha|\gamma|}\right)\right] \tag{28}$$

where $\xi_{max}$ and $\xi_0$ are two constant rheological experimental values. At every time, the values associated to the functions $\zeta_{(l)}(t)$ are obtained by solving the following equations:

$$\dot{\zeta}_{(l)}(t) + \omega_{(l)}\zeta_{(l)}(t) = \omega_{(l)}\frac{c(|\gamma|)y_{(l,0)}}{1+c(|\gamma|)\sum_{l=1}^{n}y_{(l,0)}}e(t) \tag{29}$$

where the variables $y_{(l,0)}$ are known, given by formula **(6)** for the lower strain $Q^{-1}$ value.

Finally, the rheological Eq. **(9)** is used for the considered 1D case:

$$s(t) = \frac{2G_0}{1+\alpha|\gamma|}\left[e(t) - \sum_{l=1}^{n}\zeta_{(l)}(t, y_{(l)}(\gamma))\right] \tag{30}$$

## 3   Validation of the model for cyclic loadings

The nonlinear model will be validated for 1D cyclic loadings first (homogeneous stress-strain state) directly solving Eqs (28), (29) and (30). The analysis of seismic wave propagation will be considered afterwards.

The cyclic loadings correspond to sinusoidal excitations at various strain levels. The nonlinear parameter is chosen as $\alpha=1000$ and the elastic shear modulus is $G_0=80$MPa. The relaxation parameters may then be computed considering Eqs (28) and (29) with the following asymptotic damping values: $\xi_0=0.025$ and $\xi_{max}=0.25$. In Fig. 4, some of the results (obtained at 10Hz) are displayed as stress-strain loops for $\gamma_{max}=10^{-5}$, $10^{-4}$, $5.10^{-4}$ and $10^{-3}$. For each case, the secant shear modulus $G$ is calculated and normalized by $G_0$ (the ratio $r=G/G_0$ is given in each curve).

The first case (Fig. 4, top left), corresponding to $\gamma_{max}=10^{-5}$ and $r=0.99$, leads to a nearly linear response with an elliptical stress-strain loop. In the 2$^{nd}$ case, $\gamma_{max}=10^{-4}$ and $r=0.91$ (Fig. 4, top right), the area of the loop is larger and there is a slight decrease of the shear modulus. For the largest excitations ($\gamma_{max}=5.10^{-4}$; $r=0.77$) and ($\gamma_{max}=10^{-3}$; $r=0.50$) (Fig. 4, bottom), the nonlinear effects are obvious since the stress-strain loops are strongly modified (secant modulus, area, etc).

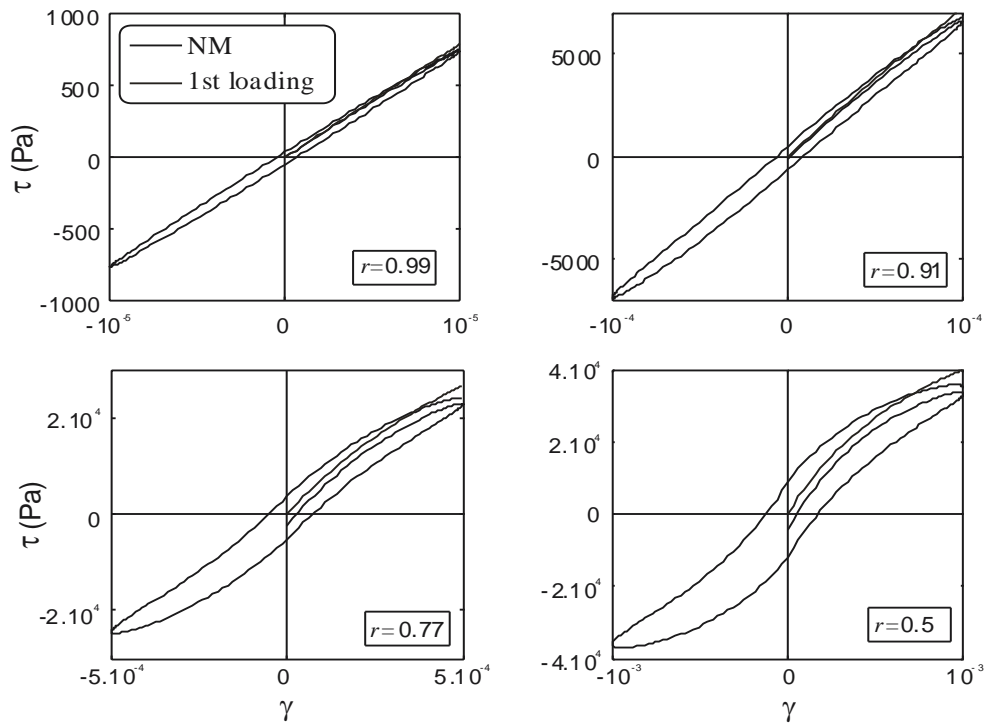

**Fig. 4.** Stress-strain curves from cyclic loadings of variable maximum amplitudes (r=G/G$_0$) at 10Hz: nonlinear extended NCQ model (solid) and 1$^{st}$ loading curve (dashed).

From these loops, it is straightforward to derive the secant shear modulus as a function of maximum shear strain. For each loading level, the dissipation may also be quantified by calculating the ratio between the area of the stress-strain loop and the strain energy estimated from the first loading curve (up to the maximum shear strain $\gamma_{max}$). The damping ratio $\xi$ may be easily derived from this energy ratio as a function of maximum shear strain (Kramer, 1996).

The actual $G(\gamma_{max})$ and $\xi(\gamma_{max})$ curves are then compared to the theoretical curves in Fig. 5. The effective shear modulus (solid) is very close from the theoretical one (dotted). For the damping ratio, the difference is larger for large shear strains, but the effective dissipation increases as expected.

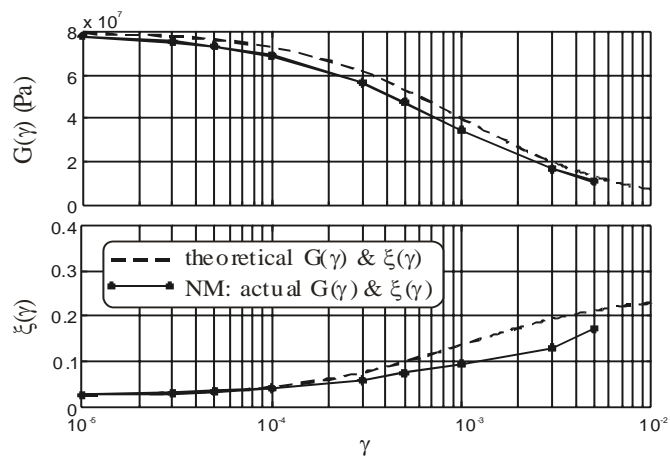

**Fig. 5.** Comparison of the shear modulus and damping values (%) of the extended NCQ model under cyclic loadings (solid) with the theoretical variations predicted by Eqs. (18) and (27) (dashed).

# Numerical implementation (FEM)

The mechanical model described above is introduced into the framework of the Finite element method, for the case of a unidirectional shear loading. Let us consider a homogeneous layer over an elastic bedrock as depicted in Fig. 6.

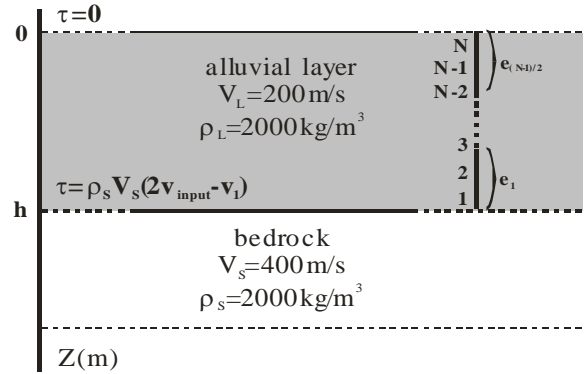

**Fig. 6.** 1D soil layer over an elastic bedrock: finite element discretization and absorbing boundary condition at the interface.

The domain is divided into *(N-1)/2* linear quadratic finite elements, each of the *N* nodes having 1 degree of freedom (horizontal motion). Using square brackets […] and braces {…} to denote matrices and vectors, the discretized equation of motion can be written in the following form at each time step *(n+1)Δt*:

$$\begin{cases} [M]\{a_{n+1}\} + [C]\{v_{n+1}\} + [K(u_{n+1})]\{u_{n+1}\} = \{F_{n+1}\} \\ \dot{\zeta}_{(l)}(t) + \omega_{(l)}\zeta_{(l)}(t) = H_{(l)}(u_{n+1}) \quad ; \quad l = 1, l_{max} \end{cases} \quad (31)$$

where $[M]$, $[C]$ and $[K(u_{n+1})]$ represent the mass, the radiation condition at the bedrock/layer interface (elastic substratum), and the stiffness matrix respectively. $\{a_{n+1}\}$, $\{v_{n+1}\}$ and $\{u_{n+1}\}$ are the acceleration, velocity and displacement vector respectively, while $\{F_{n+1}\}$ is the vector of external forces at the interface. $\zeta_{(l)}$ and $\omega_{(l)}$ are the relaxation parameters and central frequencies of the rheological cells (resp.), $H_{(l)}(u_{n+1})$ corresponds to the right hand-side term in Eq. **(29)** and $l_{max}$ is the total number of cells included in the model ($l_{max}$=3 herein).

For the time integration, an extension of the Newmark formulation is used, namely an unconditionally stable implicit *α*-HHT scheme (Hughes, 1987). This scheme allows a control of the higher frequencies generated during the propagation (Semblat and Pecker, 2009). At each time step, the Newton-Raphson iterative algorithm is adopted to deal with the nonlinear nature of the first equation in system **(31)**. The Crank-Nicolson procedure (Zienkewicz, 2005) is simultaneously used in order to estimate the $\zeta_{(l)}(t)$ variables in the first order differential equations (system **(31)**, bottom).

# 4 Modeling wave propagation in the nonlinear range

## 4.1 Nonlinear layered model

We performed two different types of simulations: linear attenuating model (denoted "LM") and nonlinear extended NCQ model (denoted "NM"). For the first one ($\xi_0 = \xi_{max} = 2.5\%$ and $\alpha = 0$), the mechanical and dissipative properties of the material do not depend on the excitation level while, in the second case ($\xi_0 = 2.5\%$, $\xi_{max} = 25\%$ and $\alpha = 1000$), both elastic and dissipative properties are function of the induced strain as shown in Figs 3 and 5.

For both models, we performed simulations for a 20m deep soil layer over an elastic bedrock, with a velocity contrast of 2 and an absorbing condition at the bottom of the layer (Fig. 6). The excitations considered hereafter thus correspond to the incident wavefield at the top of the bedrock.

### 4.2 Sinusoidal incident wavefield

In this section, the incident wavefield is a double sine-shaped acceleration wavelet similar to that proposed by Mavroeidis and Papageorgiou (Mavroeidis and Papageorgiou, 2003; Semblat and Pecker, 2009). It is defined by the following equation:

$$a(t) = \sin(\omega_0 t)\sin\left(\frac{t}{\omega_0}\right) \text{ with } \omega_0 = 2\pi f_0 \text{ and } f_0 = 3Hz \tag{32}$$

The total duration of the resulting signal is about 2 seconds.

In Fig. 7, taking into account the velocity contrast, a comparison is shown in terms of acceleration time histories and corresponding Fourier spectra at the top of the soil layer for two excitation levels (0.5 and 0.75 m/s$^2$). The nonlinear time histories involve propagation time delays when compared to the linear ones, as it can be easily observed by comparing the peaks arrival times for both models in Fig. 7. In the latter case, the Fourier spectra of the nonlinear signals indicate:

1) a significant decrease of the spectral amplitude, with increasing excitation level, for the main frequency components of the input signal;
2) the generation of higher frequency peaks which are not contained in the input signal (around 3 and 5 times the predominant excitation frequency). Such higher frequency components are larger for stronger excitations (bottom) ;
3) a frequency shift of the largest peaks to lower frequencies for increasing excitations.

The shear strain at the center of the layer is also plotted in Fig. 8 (left) for both excitation levels. Similar time delays are observed in the time-histories. From the stress-strain paths (Fig. 8, right), the reduction of the shear modulus and the energy dissipation are found to be larger for peaks of increasing amplitudes. The largest effect is obtained for the strongest excitation (Fig. 8, bottom right).

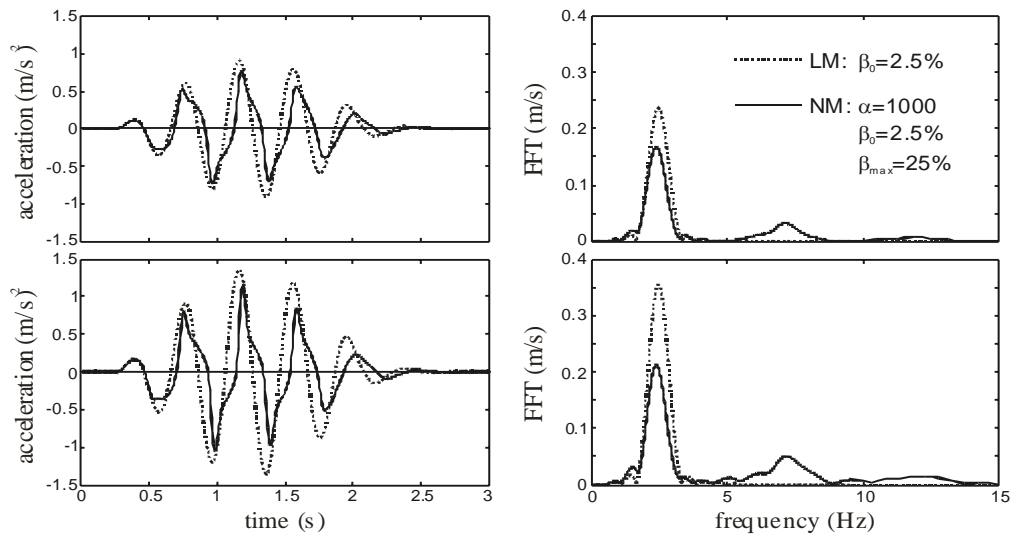

**Fig. 7.** Accelerations (left) and corresponding Fourier spectra (right) at the top of the soil layer, for 2 values of the maximum input acceleration on bedrock 0.5 (top) and 0.75 m/s$^2$ (bottom): linear (LM, dotted) and nonlinear (NM, solid) simulations.

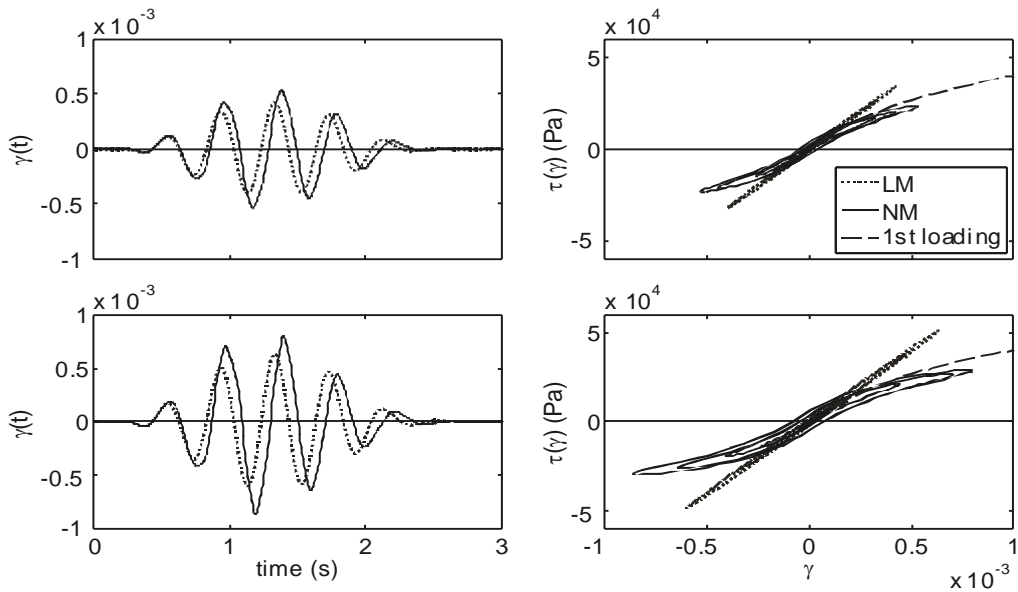

**Fig. 8.** Shear strains (left) and stress-strain loops (right) at the middle of the soil layer, for 2 values of the maximum input acceleration on bedrock 0.5 (top) and 0.75 m/s$^2$ (bottom): linear (LM, dotted) and nonlinear (NM, solid) simulations.

### 4.3 Real seismic input

#### 4.3.1 Linear and nonlinear simulations

We use the same model as in the previous case (Fig. 6) but the incident wavefield now corresponds to the horizontal acceleration recorded at Topanga station during the 1994 M6.7 Northridge earthquake (Fig. 9, top). In the linear case, the results are displayed in terms of time history and Fourier spectrum in Fig. 9 (2nd line). For the nonlinear case, two different values of the nonlinear parameter are chosen: $\alpha=300$ (Fig. 9, 3rd line) and $\alpha=600$ (Fig. 9, bottom). From the results of the linear case (2$^{nd}$ line), the incident wavefield is found to be significantly amplified at the free surface in terms of Peak Ground Acceleration (30%). Comparing the linear and the nonlinear responses, peak amplitudes in the time histories and the spectra appear to be modified. The results of the nonlinear cases lead to lower amplitudes at intermediate frequencies, whereas nonlinear responses at higher frequencies are generally larger (Fig. 9, right). It is nevertheless difficult to assess the influence of the nonlinearities for each individual peak. A time-frequency analysis is thus proposed in the next section.

In the case of the seismic excitation, the stress-strain loops are plotted in Fig. 10 for the linear and nonlinear models. When compared to the linear case (Fig. 10 left), the nonlinear cases (Fig. 10 center and right) lead to a strong modulus decrease and a large dissipation increase. The difference between both $\alpha$ values is also significant (e.g. larger loops) showing stronger nonlinear effects for the largest $\alpha$ value (*r*=0.27 for $\alpha=600$ and *r*=0.47 for $\alpha=300$).

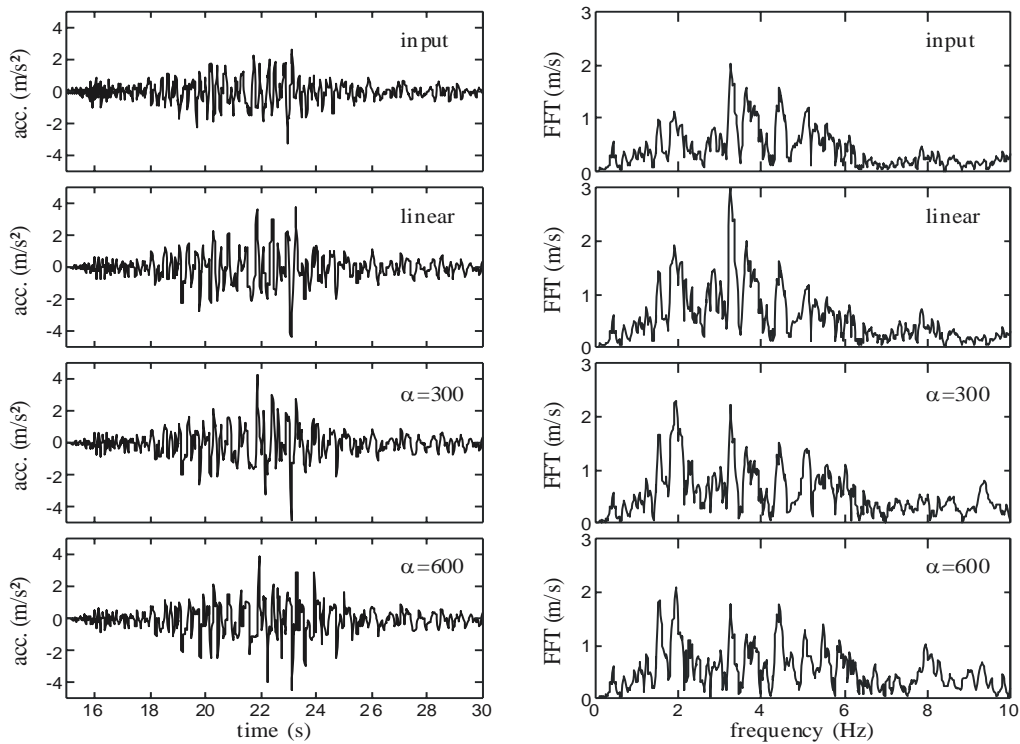

**Fig. 9.** Accelerations at the free surface for the M6.7 Northridge earthquake: time-histories (left) and related spectra (right); measured signal at Topanga station (top), linear simulation (2nd line) and nonlinear simulations with $\alpha=300$ (3rd line) and $\alpha=600$ (bottom).

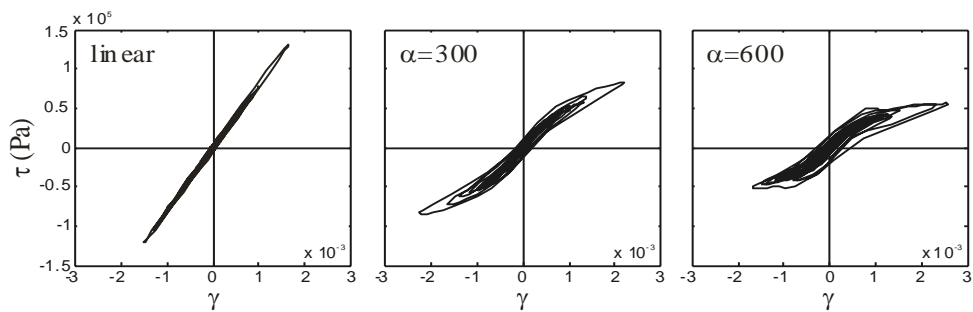

**Fig. 10.** Stress-strain curves at the middle of the soil layer for the M6.7 Northridge earthquake: linear case (left) and nonlinear cases with $\alpha=300$ (center) and $\alpha=600$ (right).

*Time-frequency analysis*

The analysis will now be performed in different frequency bands as defined in Fig. 11. In this figure, the spectral amplitudes are found to be similar for the linear and nonlinear cases in frequency bands (a) and (c), whereas bands (b) and (d) evidence significant differences. These frequency bands are the following: (a) [0-2.5Hz], (b) [2.5-4.3Hz], (c) [4.3-6.3Hz] and (d) [6.3-20Hz]. The time-histories have been (Butterworth-) filtered in each frequency band to make the comparison between the linear and nonlinear cases easier.

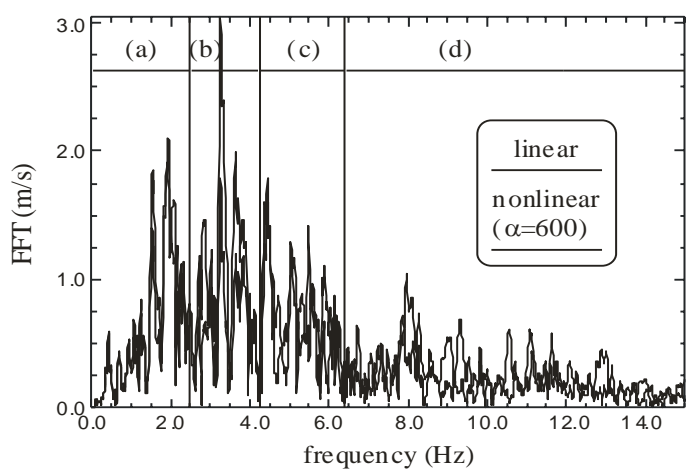

**Fig. 11.** Fourier spectra of the accelerations at the top of the soil layer (Fig. 9) for the M6.7 Northridge earthquake in the case of linear (LM, dotted) and nonlinear simulations ($\alpha$=600, solid).

The filtered accelerograms related to each frequency band are displayed in Fig. 12. The filtered linear time-histories are plotted on the left whereas the nonlinear ones ($\alpha$=600) are located on the right part. The comparison of the filtered accelerograms lead to the following conclusions:

1) *Frequency bands (a) and (c)*: the peak amplitudes of the filtered time-histories in the linear and nonlinear cases are similar. It may also be noticed in the spectra plotted in Fig. 11.
2) *Frequency band (b)*: the discrepancy between both time-histories is large since the linear response may be 30% larger than the nonlinear one. Such a difference may be directly seen in the spectra (Fig. 11).
3) *Frequency band (d)*: the nonlinear response is now larger than the linear one (up to 40%) due to the influence of higher order harmonics generated by nonlinear models (Van Den Abeele, 2000).

For strong seismic motion, the nonlinear ground response may then be smaller or larger than the linear one depending on the excitation level as well as the frequency content of the input motion. The nonlinear properties of the soil are also an important governing parameter of its seismic response.

## 5 Conclusions

A 3D nonlinear viscoelastic model ("extended NCQ") is proposed to approximate the hysteretic behavior of alluvial deposits undergoing seismic excitations. Such nonlinear features as the reduction of shear modulus and the increase of damping are controlled by the variations of the $2^{nd}$ invariant of the strain tensor during multidimensional loading. In the case of a unidirectional shear loading, nonlinearity is controlled by only one shear strain component: nonlinear elasticity by a hyperbolic law and viscosity by a NCQ model with nonlinear features (nearly frequency constant but strain amplitude dependent).

This model allows to account for the generation of higher order harmonics shown in the nonlinear case for 1D simulations. At the same time, a reduction of the spectral amplitudes and a shift to lower frequencies were found for increasing motion amplitudes. The interest of the simplified nonlinear "X-NCQ" model proposed herein is to reduce the computational cost for the analysis of strong seismic motion in 2D/3D alluvial basins (small number of constitutive parameters).

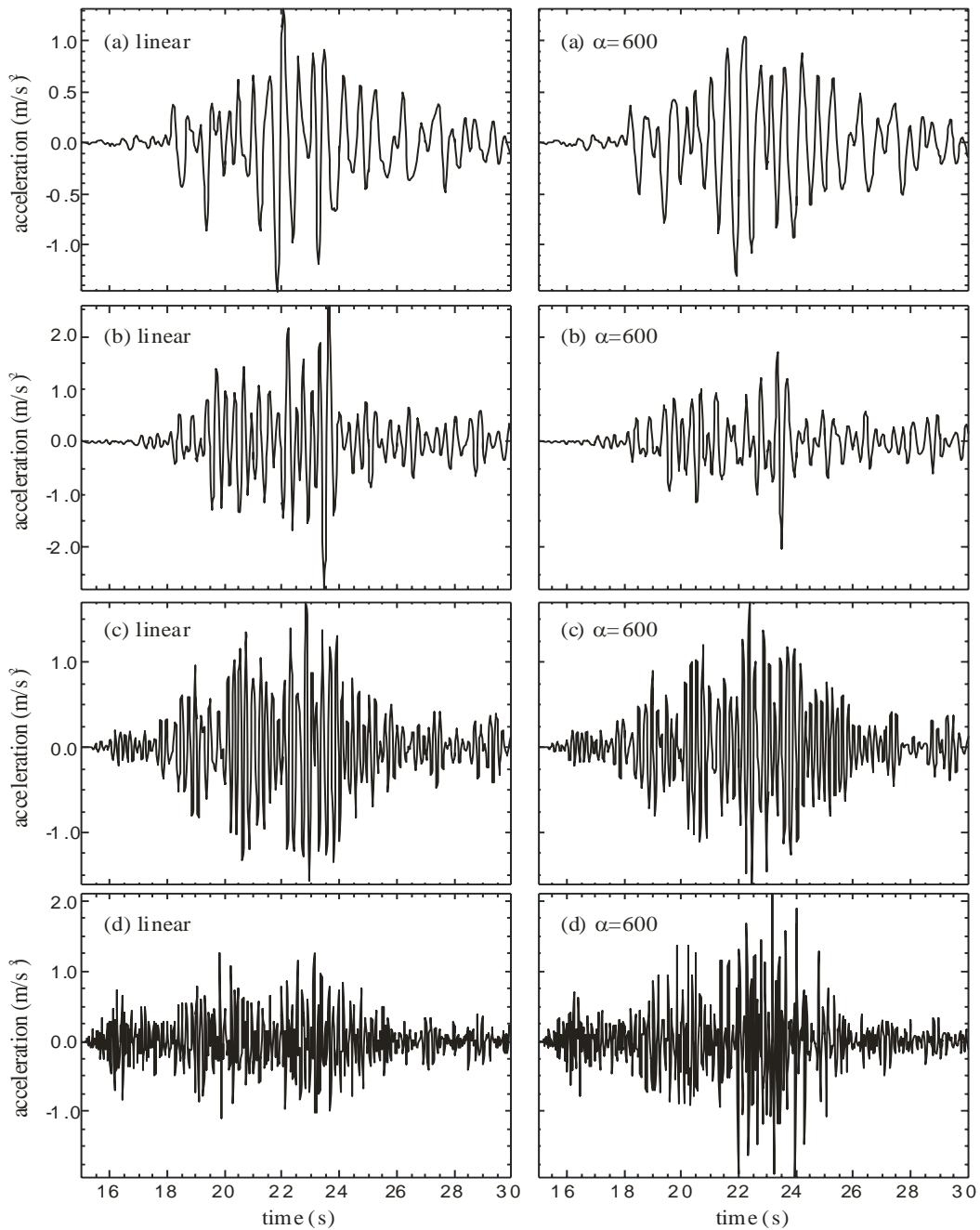

**Fig. 12.** Accelerations at the top of the soil layer for the M6.7 Northridge earthquake in the case of linear (LM, dotted) and nonlinear simulations ($\alpha$=600, solid) filtered in different frequency bands defined in Fig. 11.

For example, in the 1D case, the reduction of shear modulus is controlled by a hyperbolic law with only one parameter $\alpha$ estimated from the experimental knowledge of the $G(\gamma)$ curve. As a consequence, the dissipation properties are directly derived from the hyperbolic law and from two other characteristic parameters responsible for the minimum and maximum loss of energy at lower and larger strain levels, $\xi_0$ and $\xi_{max}$. These are sufficient to give an overall description of the unloading and reloading phases during the seismic sequence. Combined with the nonlinear properties of the soil in the simplified model, the frequency content of the seismic input has an important influence on strong ground motions. Finally, the proposed model will allow future computations in the case of 2D or either 3D alluvial basins for which the amplification is generally found to be much larger than predicted through 1D analyses (Chaillat et al., 2009; Chávez-García et al., 1999; Fäh et al., 1994; Gélis et al., 2008; Lenti et al., 2009;

Moeen-Vaziri and Trifunac, 1988; Sánchez-Sesma and Luzón, 1995; Semblat et al., 2000, 2005). Several authors proposed some 2D/1D aggravation factors (Makra et al., 2005; Semblat and Pecker, 2009), but it is probably not sufficient for strong seismic motions involving significant nonlinerities in the soil response.

APPENDIX:
Emmerich and Korn's method to find the optimal parameters of the linear viscoelastic "NCQ" model is presented in this appendix.

We consider the viscoelastic model depicted in Fig. 1 (left). To estimate the $a_{(l)}$ coefficients, a normalization condition is introduced:

$$y_{(l,0)} = \frac{\delta M}{M_R} a_{(l)} \tag{33}$$

The $\omega_{(l)}/\omega_{(l-1)}$ ratio being chosen constant, Eq. **(6)** is simplified as:

$$Q^{-1}(\omega) \approx \sum_{l=1}^{n} y_{(l,0)} \frac{\omega/\omega_{(l)}}{1+\left(\omega/\omega_{(l)}\right)^2} \tag{34}$$

The $y_{(l,0)}$ quantities are estimated by using Eq. (34): writing it for different $\omega$ and for several fixed values of $\omega_{(l)}$ and taking the first term equal to a given constant value, the obtained algebraic linear system can be solved by a least-squares algorithm. An example of the result of this procedure is displayed in Fig. 2: in the case of $\xi$=2.5% (Q=20) and a velocity of 200m/s. A normalization condition allows to choose a target phase velocity (200m/s) at a given reference frequency (1Hz in the example).

For more details, the readers may refer to Emmerich and Korn (1987).

**Acknowledgements:**
The authors would like to thank Luis F. Bonilla (IRSN) for fruitful discussions. This work was partly funded by the French National Research Agency in the framework of the "QSHA" research project ("Quantitative Seismic Hazard Assessment").

**Notations:**
The following symbols are used in this paper:

| | | |
|---|---|---|
| $\{a\}$ | = | acceleration vector (FEM) |
| $[C]$ | = | damping matrix (FEM) |
| $c(\|\gamma_{oct}\|)$ | = | weighting function for non linear damping |
| *e* | = | shear deviatoric strain tensor |
| $e_{ij}(\omega)$ | = | Fourier transforms of the components of the deviatoric strain |
| $e_{kk}$ | = | volumetric strain |
| $\{F\}$ | = | external force vector (FEM) |
| *f* | = | frequency |
| $G_0$ | = | (unrelaxed) shear modulus at low strains |
| $I'_1$ | = | first invariant of the strain tensor |
| $I'_2$ | = | second invariant of the strain tensor |
| $J_2$ | = | second invariant of the deviatoric strain tensor |
| $K$ | = | bulk modulus |
| $[K]$ | = | tangent stiffness matrix (FEM) |
| $M(\omega)$ | = | complex viscoelastic modulus |
| $[M]$ | = | mass matrix (FEM) |
| $M_R$ | = | relaxed modulus |
| $M_U$ | = | unrelaxed modulus |
| $M_{U,0}$ | = | unrelaxed modulus at low strains |
| $p$ | = | volumetric tension |
| $Q$ | = | quality factor |
| $Q^{-1}$ | = | specific attenuation |
| $s$ | = | shear deviatoric stress tensor |
| $s_{ij}$ | = | components of the deviatoric stress tensor |
| $s_{ij}(\omega)$ | = | Fourier transforms of the components of the deviatoric stress |
| $\{u\}$ | = | displacement vector (FEM) |
| $\{v\}$ | = | velocity vector (FEM) |
| $y_{(l,0)}$ | = | relaxation parameters of the viscoelastic cells for low excitation levels |
| $\alpha$ | = | scalar parameter characterizing the modulus reduction |
| $\gamma_{oct}$ | = | octahedral strain |
| $\delta_{ij}$ | = | Kronecker unit tensor components |
| $\delta M$ | = | difference between the relaxed and unrelaxed moduli |
| $\zeta_l(t)$ | = | relaxation functions |
| $\xi_0$ | = | minimum damping at low strains |
| $\xi_{max}$ | = | maximum damping at large strains |
| $\sigma_{ij}$ | = | components of the Cauchy stress tensor |
| $\Phi$ | = | function characterizing the modulus reduction |
| $\omega$ | = | circular frequency |